\documentclass[twocolumn,showpacs,pra,preprintnumbers,superscriptaddress,amsmath,amssymb]{revtex4}
\usepackage{bbold}
\usepackage{color}
\usepackage{amssymb}
\usepackage{graphicx}
\usepackage{dcolumn}
\usepackage{bm}
\usepackage{amsmath}
\newcommand{\be}{\begin{eqnarray}}
\newcommand{\ee}{\end{eqnarray}}

\begin{document}
\title{Fate of Weyl semimetals in the presence of incommensurate potentials}
\author{Yucheng Wang}
\affiliation{Beijing National
Laboratory for Condensed Matter Physics, Institute of Physics,
Chinese Academy of Sciences, Beijing 100190, China}
\author{Shu Chen}
\thanks{schen@iphy.ac.cn}
\affiliation{Beijing National
Laboratory for Condensed Matter Physics, Institute of Physics,
Chinese Academy of Sciences, Beijing 100190, China}
\affiliation{School of Physical Sciences, University of Chinese Academy of Sciences, Beijing, 100049, China}
\affiliation{Collaborative Innovation Center of Quantum Matter, Beijing, China}
\begin{abstract}
We investigate the effect of the incommensurate potential on Weyl semimetal, which is proposed to be realized in  ultracold atomic systems trapped in three-dimensional optical lattices. For the system without the Fermi arc, we find that the Weyl points are robust against the incommensurate potential and the system enters into a metallic phase only when the incommensurate potential strength exceeds a critical value. We unveil the trastition by analysing the properties of wave functions and the density of states as a function of the incommensurate potential strength. We also study the system with Fermi arcs and find the Fermi arcs are sensitive against the incommensurate potential  and can be destoryed by a weak incommensurate potential.
\end{abstract}
\maketitle
\section{Introduction}
During the past few years, considerable attention has been paid on topological phases of matters, which include the gapped topological insulators \cite{Kane,Zhang} and various gapless systems \cite{Nagaosa,Fang}. Weyl semimetal (WSM), as a typical example of topologically nontrivial gapless systems, has been widely studied in both theories \cite{Murakami,Wan,Burkov,Yang,Xu,Halasz,Young,Huang} and experiments \cite{Lv,Hasan,Hasan2,Yang2}.   A WSM is a three-dimensional topological semimetal which has some isolated touching points between conduction and valence bands. These touching points, named as Weyl nodes, have definite chiralities and can be understood as topologically protected chiral charges. Weyl nodes with opposite chiralities in momentum space can be connected by nonclosed surface states, known as Fermi arc states \cite{Wan}.

The effects of disorder on Weyl semimetals \cite{Huang2,Ominato,Brouwer,Ryu,Xie,Hughes,Roy,Gurarie} and Dirac semimetals \cite{Herbut,Sondhi,Sarma1,Sarma2} have been a subject of intensive study. It has been found that a WSM phase or a Dirac semimetal phase is robust against weak disorder and there exist semimetal-metal-insulator quantum phase transitions when the strength of disorder increases. Some recent works proposed to realize the Weyl Hamiltonian for ultracold atoms in three-dimensional optical lattices by using laser-assisted tunneling \cite{Buljan,Lan,Jiang,Delplace,Dan,Law,XuY,Fulga}. Besides, cold atoms in optical lattices have been widely used to simulate various models of topological insulators \cite{Stanescu,Goldman,Goldman2,Cooper,Goldman3}. Particularly, by manipulating the atomic
hopping configurations in optical lattices, the famous Harper-Hofstadter \cite{Harper} and Haldane models \cite{Haldane} have already been experimentally
realized \cite{Bloch,Jotzu,Aidelsburger}. On the other hand, one can generate incommensurate optical lattices by superimposing two one-dimensional (1D) optical lattices with incommensurate wavelengths \cite{Roati}, which has been widely applied to study the localization to delocalization transition induced by the incommensurate potential \cite{AA,Machida,Geisel,Niu,Roscilde,Albert,He,Lahini,Cai,roux,Modugno,Deissler} and the experimental exploration of many-body localization \cite{Schreiber2}. One interesting question is what is the fate of the WSM when the incommensurate potential is introduced in one direction of the three-dimensional optical lattice for a Weyl Hamiltonian. To examine this question, we add an additional incommensurate potential to the proposed Weyl Hamiltonian \cite{Buljan} and investigate the effect of incommensurate potential on the WSM phase.

This paper is organized as follows: in Sec. \ref{no}, we introduce the model of Weyl Hamiltonian with additional incommensurate potential added in one direction of the three-dimensional optical lattice. We firstly consider the case in the absence of Fermi arc and investigate the extended-localization transition of the system in the direction with incommensurate potential by studying the inverse participation ratio (IPR) of wave functions. By analysing the change of density of states (DOS) of this system, we demonstrate the occurrence of a transition from WSM to a two-dimensional metallic phase. We then consider the case in the presence of Fermi arcs and study the effect of incommensurate potential on the Fermi arcs. A brief summary is given in Sec. \ref{conclusion}.
\begin{figure}[h]
\includegraphics[height=120mm,width=80mm]{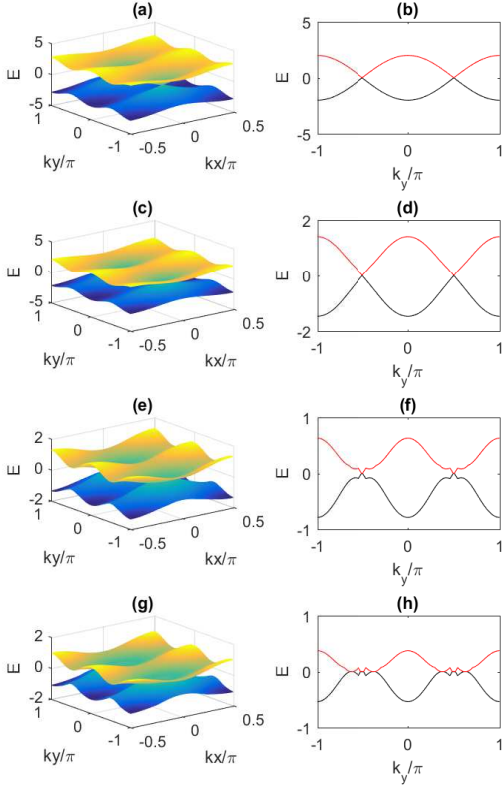}
\caption{\label{01}(Color online)
  Dispersions of the $L-th$ and $(L+1)-th$ levels of the system with (a) $V=0$, (c) $V=1$, (e) $V=2.1$ and (g) $V=2.5$ and the corresponding dispersions versus $k_y$ with fixed $k_x=0$ in (b), (d), (f) and (h), respectively. Here we consider the system with $t_x=t_y=t_z=1$ and $L=300$ under OBC in the $z$ direction and PBC in the $x$ and $y$ directions.}
\end{figure}

\section{Model and results}
\label{no}
We consider the model described by
\be
H_0 &=& -\sum_{m,n,l}(t_x e^{-i\phi_{m,n,l}}a^{\dagger}_{m+1,n,l}a_{m,n,l}+t_ya^{\dagger}_{m,n+1,l}a_{m,n,l} \nonumber\\
&+& t_ze^{-i\phi_{m,n,l}}a^{\dagger}_{m,n,l+1}a_{m,n,l}+{\rm H.c.}),
\label{ham-1}
\ee
where $a_{m,n,l}$ ($a^{\dagger}_{m,n,l}$) is the annihilation (creation) operator on the site $(m,n,l)$, $t_x$, $t_y$ and $t_z$ are the hopping strengths along $x$, $y$ and $z$ directions, respectively, and $\phi_{m,n,l}=(m+n)\pi$ (modulo $2\pi$).
In principle, this model can be realized in three-dimensional optical lattices by engineering the atomic hopping configurations as proposed in Ref. \cite{Buljan}.
After applying Bloch's theorem, one can obtain the Hamiltonian in the momentum space
\be
H_0(k) &=& -2 [ t_y \cos (k_y a) \sigma_x+t_x \sin (k_xa) \sigma_y \nonumber\\
 &-& t_z \cos (k_za) \sigma_z ],
\label{ham-k1}
\ee
where $\sigma_{x,y,z}$ are Pauli matrices, $k_x$, $k_y$ and $k_z$ are wave vectors defined in the first Brillouin of an $L_x\times L_y\times L_z$ cubic lattice, and $a$ is the lattice constant. For convenience, we shall set $a=1$ in the following context.

The energy spectrum of the Hamiltonian is given by
\begin{equation}
E(k)=\pm 2\sqrt{t_x^2 \sin^2(k_x)+t_y^2 \cos^2(k_y)+t_z^2 \cos^2(k_z)},
\label{ham-k2}
\end{equation}
which consists of two bands touching at four Weyl points $(k_x,k_y,k_z)=(0,\pm\pi/2,\pm\pi/2)$ \cite{Buljan}.
If we take the open boundary conditions (OBC) along the $z$ direction and periodic boundary conditions (PBC) in the $x$ and $y$ directions, $k_z$ is no longer a good quantum number. Nevertheless, the upper and lower bands still touch at the Weyl points $(k_x,k_y)=(0,\pm \pi/2)$. For simplicity, we consider the system with $L_x=L_y=L_z=L$. To see the spectrum clearly, we show the energy dispersions of the $L-th$ and $(L+1)-th$ levels for the system with $t_x=t_y=t_z=t$ and $L=300$ in Fig.~\ref{01}(a) and the energy as a function of $k_y$ by fixing $k_x=0$ in Fig.~\ref{01}(b). For convenience, we set $t=1$ as the energy unit. One can find that the dispersions around the two Weyl points are linear and there exists no Fermi arc. Then we add quasiperiodic potential along the $z$ direction and the Hamiltonian can be described by
\begin{equation}
\begin{aligned}
H=H_0 + V \sum_{l} \cos(2\pi\alpha l),
\label{ham-2}
\end{aligned}
\end{equation}
where $\alpha$ is an irrational number chosen as $\alpha=(\sqrt{5}-1)/2$.

Since $k_x$ and $k_y$ are good quantum numbers with $k_x=\frac{2\pi}{L}i_x$ and $k_y=\frac{2\pi}{L}i_y$, where $i_x=-\frac{L}{4},-\frac{L}{4}+1,\cdots,\frac{L}{4}-1$ and $i_y=-\frac{L}{2},-\frac{L}{2}+1,\cdots,\frac{L}{2}-1$, we can still diagonalize the Hamiltonian (\ref{ham-2}) in the momentum space of $k_x$ and $k_y$.
By representing the $n-th$ eigenstate as $|\Psi_n\rangle= [\psi_{n,1,A},\psi_{n,1,B},\psi_{n,2,A},\psi_{n,2,B},\cdots, \psi_{n,L,A},\psi_{n,L,B}]^{T}$ and using $H|\Psi_n\rangle=E_n|\Psi_n\rangle$,  one can obtain the following explicit forms:
\begin{widetext}
\begin{equation}
\begin{split}
&E_n\psi_{n,j,A}=t_z(\psi_{n,j-1,A}+\psi_{n,j+1,A})+V \cos(2\pi\alpha j)\psi_{n,j,A}+(-2 t_y \cos k_y +2i t_x \sin k_x )\psi_{n,j,B},\\
&E_n\psi_{n,j,B}=-t_z(\psi_{n,j-1,B}+\psi_{n,j+1,B})+(-2t_y \cos k_y-2it_x \sin k_x )\psi_{n,j,A}+V \cos(2\pi\alpha j)\psi_{n,j,B},
\label{ap1}
\end{split}
\end{equation}
\end{widetext}
where $E_n$ is the $n$-th eigenvalue of the system and $j$ represents the $j-th$ layer along the $\hat{z}$ direction. Given a group of values of fixed $k_x$ and $k_y$, the eigenvalue problem of the Hamiltonian (\ref{ham-2}) reduces to diagonalize a $2L\times 2L$ matrix. By solving Eq. (\ref{ap1}), we can get the energy dispersion $E_n(k_x,k_y)$ for various $V$ and study the change of energy spectrum. In Fig.~\ref{01}(c)-(h), we present dispersions of the $L-th$ and $(L+1)-th$ levels for the system with $V=1$, $2.1$ and $2.5$, respectively. In Fig.~\ref{01}(c), (e), (g), we show $E_n(k_x,k_y)$ versus $k_x$ and $k_y$, and in Fig.~\ref{01}(d), (f), (h) the corresponding $E_n$  as a function of $k_y$ with fixed $k_x=0$. While the Weyl points at $(k_x,k_y)=(0,\pm \pi/2)$ are not destroyed when the incommensurate potential strength is not so strong, e.g., $V=1$ as shown in  Fig.~\ref{01}(c) and (d), the spectrum structure is completely changed when $V=2.5$, as shown in  Fig.~\ref{01}(g) and (h).

\begin{figure}
\includegraphics[height=120mm,width=80mm]{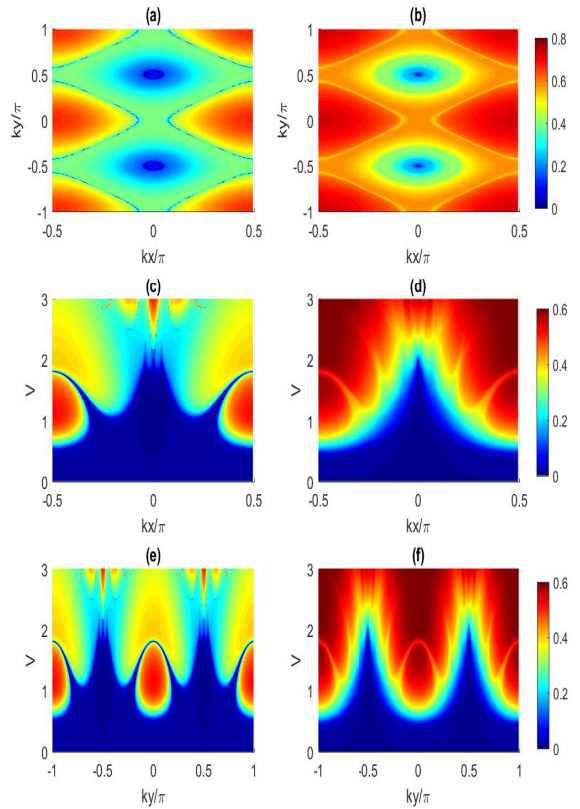}
\caption{\label{02}(Color online)
  (a) IPR and (b) MIPR as a function of $k_x$ and $k_y$ with fixed $V=1.9$. (c) IPR and (d) MIPR as a function of $k_x$ and $V$ with fixed $k_y=\frac{\pi}{2}$. (e) IPR and (f) MIPR as a function of $k_y$ and $V$ with fixed $k_x=0$. The lattice size is $L=300$ and $t_x=t_y=t_z=1$.}
\end{figure}

To investigate the extended-localization transition in the $z$ direction, we introduce the IPR \cite{Thouless,Schreiber} of the system with fixed $k_x$ and $k_y$, which is defined as
\begin{equation}
IPR = \sum_j(\psi_{n,j,A}^2+\psi_{n,j,B}^2)^2
\end{equation}
for a normalized wave function $\Psi_n$. The IPR is a useful quantity to characterize the delocalization-localization transition of a disorder or an incommensurate system. The IPR approaches to zero in the thermodynamic limit for an extended state, but tends to a finite value of $O(1)$ for a localized state. We further define the mean IPR as
\begin{equation}
MIPR=\frac{1}{2L}\sum^{2L}_{n=1}\sum_j(\psi_{n,j,A}^2+\psi_{n,j,B}^2)^2,
\end{equation}
which is the average of IPRs for all the eigenstates.
We display the IPR of the $L-th$ eigenstate in Fig.~\ref{02}(a) and the MIPR in Fig.~\ref{02}(b) as a function of $k_x$ and $k_y$ for $V=1.9$.  It is shown that states around $(k_x,k_y)=(0,\pm \pi/2)$ are still in extended states, whereas states in the corner and side regimes of momentum space already become localized in the $z$ direction, i.e., states around $(k_x,k_y)=(0,\pm \pi/2)$ are harder to become localized than states in the other regimes. In order to localize the states around the Weyl points, one needs to increase the incommensurate potential strength $V$. This property can be further clarified from Fig.~\ref{02}(c) and Fig.~\ref{02}(d), which show the IPR of the $L-th$ eigenstate and MIPR as a function of $k_x$ and $V$ with the fixed $k_y=\frac{\pi}{2}$, respectively. Similarly, in Fig.~\ref{02}(e) and Fig.~\ref{02}(f), we show the IPR and MIPR as a function of $k_y$ and $V$ by fixing $k_x=0$. From these figures, we see that some eigenstates become localized along the $z$ direction when $V$ increases over about $0.4$, however states around $(k_x,k_y)=(0,\pm \pi/2)$ become localized when $V>2$.
This phenomenon can be intuitively understood by observing Eq. (\ref{ap1}). Given fixed $k_x$ and $k_y$, Eq. (\ref{ap1}) can be viewed as describing an effective 1D tight-binding model with the incommensurate potential. The effective hopping strengths between the A and B sublattices depend on the values of  $k_x$ and $k_y$. When $k_x=0$ and $k_y=\pm \frac{\pi}{2}$, the effective hopping strengths between two sublattices are zero, and the model can be mapped to two decoupled 1D Aubry-Andr\'{e} (AA) models \cite{AA}. It is well known that all single particle states in the AA model are extended when $V<2$, whereas they are localized when $V>2$. This is consistent with our result that all eigenstates are localized along the $z$ direction when $V>2$.
\begin{figure}
\includegraphics[height=50mm,width=85mm]{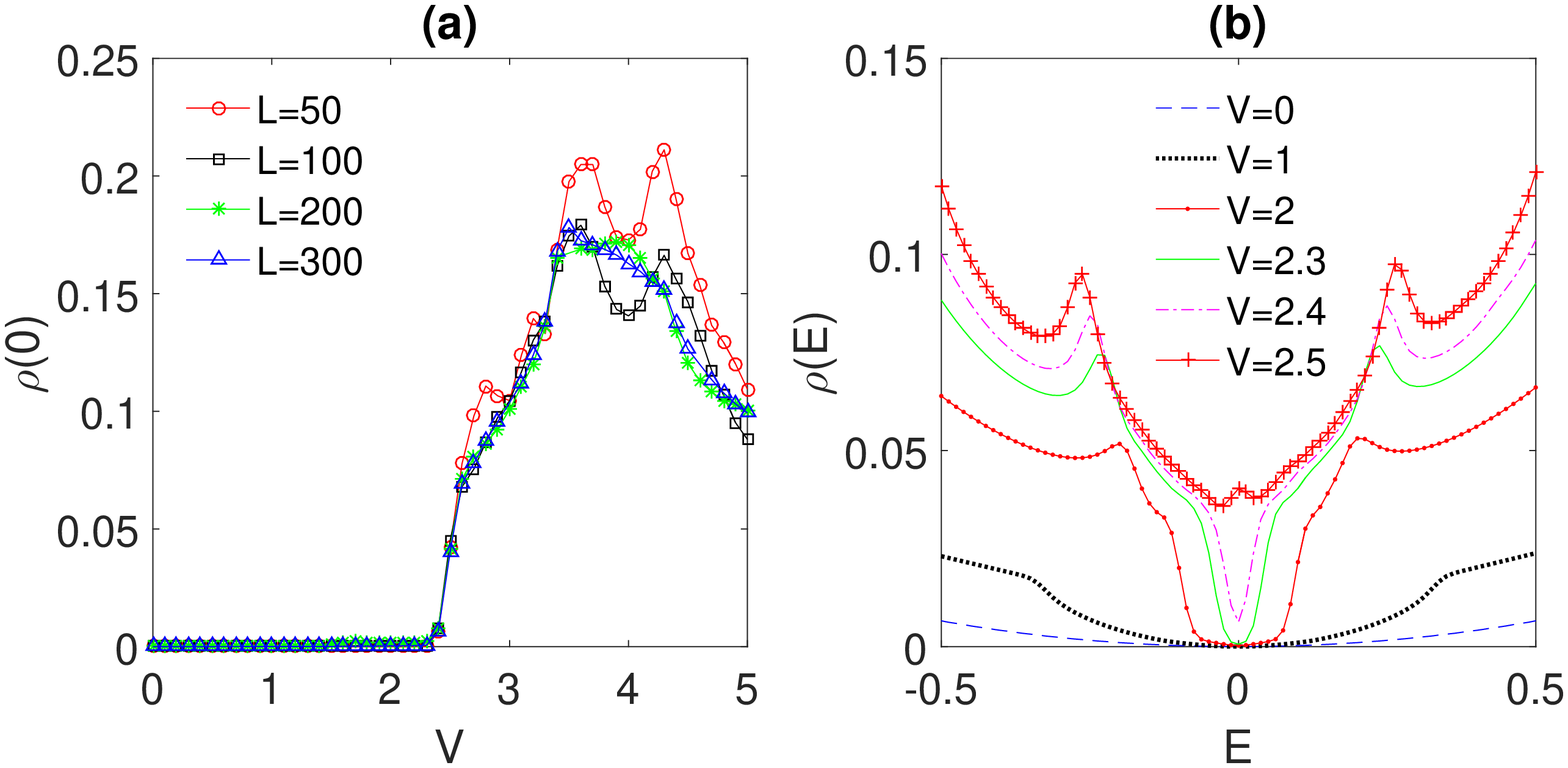}
\caption{\label{03}(Color online)
  (a) $\rho(0)$ versus $V$ for different lattice size $L$ with fixed $\sigma=0.02$. (b) DOS with $L=300$ as a function of energy for various values of incommensurate potential strength $V$. Here $t_x=t_y=t_z=1$.}
\end{figure}

To further see how the WSM is changed against the increasing of incommensurate potential, we calculate the DOS, which is defined as
\be
\rho(E) &=& \frac{1}{L^3}\sum_{i=1}^{L^3}\delta(E-E_i)\nonumber\\
&=& \frac{1}{L^3}\sum_{l=1}^{2L}\sum_{i_x=-\frac{L}{4}}^{\frac{L}{4}-1}\sum_{i_y=-\frac{L}{2}}^{\frac{L}{2}-1}\delta(E-E_{l,i_x,i_y}).
\label{dos}
\ee
Here $E_{l,i_x,i_y}$ is the $l-th$ eigenstate with fixed $k_x=\frac{2\pi}{L}i_x$ and $k_y=\frac{2\pi}{L}i_y$. To numerically calculate $\delta(E-E_i)$, we make an approximation by replacing the function of $\delta(x)$ by a Gaussian function
$\frac{1}{\sqrt{\pi\sigma^2}}exp(-\frac{x^2}{\sigma^2})$ \cite{Fehske},
which approaches the $\delta$-function exactly when $\sigma \rightarrow 0$.
The WSM is characterized by the DOS $\rho(E)\sim |E|^2$ \cite{Roy}, which gives rise to $\rho(0)=0$. On the other hand, the DOS $\rho(0)$ becomes finite if the system enters into a metallic phase. So $\rho(0)$ can give signatures of the transition from a WSM to metal phase. Fig.~\ref{03}(a) gives the DOS at zero energy as a function of incommensurate potential strength for different $L$. It is shown that the transition point isn't sensitive to the lattice size when $L>50$, so it is reliable to choose $L=300$. From Fig.~\ref{03}(a), the WSM-metal transition point can be found at about $V=2.3$. Fig.~\ref{03}(b) shows the DOS with $L=300$ as a function of energy for various values of $V$. For $V=0$, $\rho(E)\sim |E|^2$. This relation holds true in the region of $|E| \sim 0$  even in the presence of a finite $V$ and we have $\rho(0) =0$  as long as $V$ is less than $2.3$. Such a relation no longer holds true when $V$ is larger than $V_c=2.3$, as $\rho(0)$ rapidly increases and becomes finite once $V>V_c$.

\begin{figure}
\includegraphics[height=60mm,width=80mm]{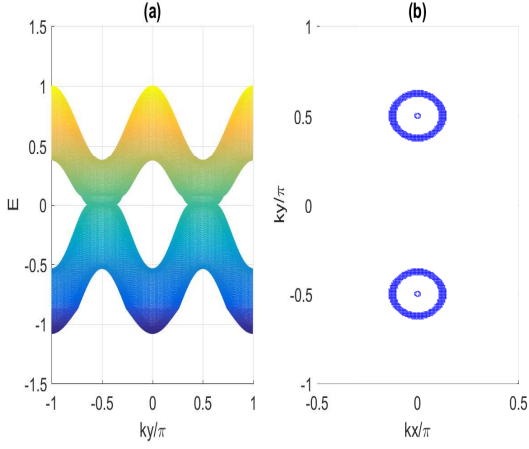}
\caption{\label{04}(Color online)
  (a) The projection of dispersions of $L-th$ and $(L+1)-th$ levels of the system with $V=2.5$ and $t_x=t_y=t_z=1$ on the $k_y-E$ plane. (b) The cross section of energy dispersion $E(k_x,k_y)$ at $E=0\pm 0.001$.}
\end{figure}

The change of DOS is closely related to the change of spectrum structure of the system. We find that the structure of the energy dispersion $E(k_x,k_y)$ does not change dramatically at $V=2$, even all eigenstates become localized in the $z$ direction when $V>2$. For example, as shown in Fig.~\ref{01}(e) and (f), for $V=2.1$, the Weyl points don't disappear and the dispersion relation is still linear near the Weyl points. However, when increasing $V$ further to exceed $V_c$, for example $V=2.5$ as shown in Fig.~\ref{01}(g) and Fig.~\ref{01}(h), one can find the touched segment of the $L-th$ and $(L+1)-th$ levels of this system no longer has a similar structure of Weyl points. To be clear, we show the spectrums of $L-th$ and $(L+1)-th$ levels projected onto the $k_y-E$ plane in Fig.~\ref{04}(a) and the cross section at $E=0\pm 0.001$ (with considering the size effect) in Fig.~\ref{04}(b). It is clear that the two-band touched segment is no longer composed of some single points and there exist many states for $E=0$, i.e., $\rho(0)$ becomes a finite value. Correspondingly, the system becomes a metal. In this metal phase, we note that the wave functions in the $z$ direction are localized but the wave functions in  the $x$ and $y$ directions are extended, i.e., the system can be viewed as a two-dimensional metal.


If we take OBC in the $\hat{x}-\hat{y}$ direction and PBC in the $\hat{z}$ and $\hat{x}+\hat{y}$ directions, $k_{||}(\hat{x}+\hat{y})$ and $k_z$ are good quantum numbers and the Weyl points on the $(k_{||},k_z)$ plane are at $(k_{||},k_z)=(\pm \frac{\pi}{2\sqrt{2}},\pm \frac{\pi}{2})$, which are connected with Fermi arcs \cite{Buljan}. We set $L_{||}=L_{\hat{x}-\hat{y}}=L=300$ and add incommensurate potential $V\cos(2\pi\alpha m)$ along the $\hat{x}-\hat{y}$ direction, where $m$ represents the $m-th$ layer along the $\hat{x}-\hat{y}$ direction. By using a similar method as the case in the absence of Fermi arc, we can obtain the following eigen-equations:
\begin{widetext}
\begin{equation}
\begin{split}
E_n\psi_{n,j,A}=&(-t_ye^{ik_{||}/\sqrt{2}}-t_xe^{-ik_{||}/\sqrt{2}})\psi_{n,j-1,B}+[2t_z\cos k_z+V \cos(2\pi\alpha j)]\psi_{n,j,A}\\ &+(-t_ye^{-ik_{||}/\sqrt{2}}+t_xe^{ik_{||}/\sqrt{2}})\psi_{n,j,B},\\
E_n\psi_{n,j,B}=&(-t_ye^{-ik_{||}/\sqrt{2}}-t_xe^{ik_{||}/\sqrt{2}})\psi_{n,j+1,A}+(-t_ye^{ik_{||}/\sqrt{2}}+t_xe^{-ik_{||}/\sqrt{2}})\psi_{n,j,A}\\
&+ [ -2t_z\cos k_z + V \cos(2\pi\alpha j)]\psi_{n,j,B},
\label{ap2}
\end{split}
\end{equation}
\end{widetext}
where $k_{||}$ belongs to $[-\pi/\sqrt{2},\pi/\sqrt{2})$, $k_z$ belongs to $[-\pi,\pi)$ and $j$ represents the $j-th$ layer along the $\hat{x}-\hat{y}$ direction.

\begin{figure}
\includegraphics[height=90mm,width=80mm]{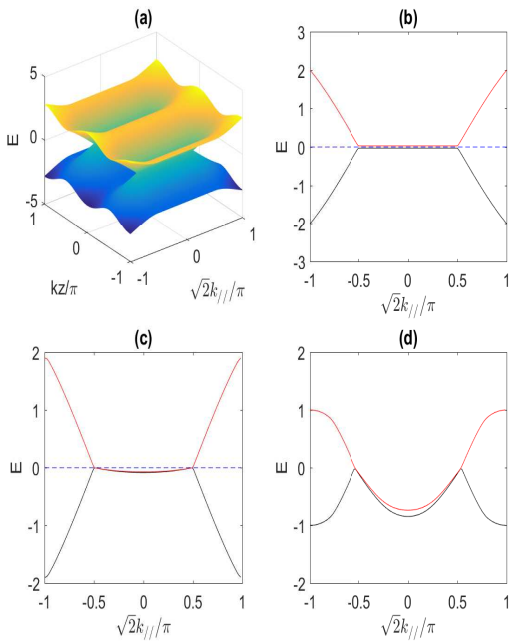}
\caption{\label{05}(Color online)
  (a) Dispersions of the $L-th$ and $(L+1)-th$ levels of the system with $V=0$. (b)-(d) Dispersions of the $L-th$ and $(L+1)-th$ levels as a function of $k_{||}$ with fixed $k_z=\frac{\pi}{2}$ for the system with (b) $V=0$, (c) $V=0.1$ and (d) $V=1$. Here we have taken $t_x=t_y=t_z=1$. The dotted lines of (b) and (c) are the referenced line of $E=0$.}
\end{figure}
In Fig.~\ref{05}(a), we display the energy dispersion of the $L-th$ and $(L+1)-th$ levels of the system with $t_x=t_y=t_z=1$ and $V=0$, and in Fig.~\ref{05}(b) the spectrum as a function of $k_{||}$ by fixing $k_z=\frac{\pi}{2}$. It is shown that there exists Fermi arc connecting the Weyl points at $k_{||}=\pm \frac{\pi}{2\sqrt{2}}$ in the absence of incommensurate potential. When the incommensurate potential with the strength $V=0.1$ is added, the dispersion shapes of the $L-th$ and $(L+1)-th$ levels change. As shown in Fig.~\ref{05}(c), the Fermi arc corresponding to the curve of $E=0$ is destroyed, and the Weyl points are connected by curves of $E(k_{||},k_z)<0$. For a half-filled system, the Fermi surface $E_c$ moves down to $E_c<0$, and the system enters into a metallic phase. Further increasing the incommensurate potential strength $V$, the dispersion shape of the $L-th$ and $(L+1)-th$ levels change more dramatically as shown in Fig.~\ref{05}(d), and there exists a bigger region with the $(L+1)-th$ level entering into the region of $E<0$. Our results indicate that the WSM phase for the case without Fermi arcs is more robust against to the incommensurate potential than the case in the presence of Fermi arcs.

\section{Summary}
\label{conclusion}
In summary, we have studied the effect of incommensurate potential on WSM either in the absence or the presence of Fermi arcs. We show that the incommensurate potential plays obviously different roles in these two cases. For the system without Fermi arcs, the WSM  is robust against the incommensurate potential. By calculating the DOS of the system, we show that the system enters into a metallic phase when the incommensurate potential strength exceeds a critical value, which is even bigger than the extended-localization transition point along the $z$ direction.  However, for the system in the presence of Fermi arcs, we show that the Fermi arcs are sensitive against the incommensurate potential, and the WSM is unstable even for a very small incommensurate potential.

\begin{acknowledgments}
The work is supported by the National Key Research and Development Program of China (2016YFA0300600), NSFC under Grants No. 11425419, No. 11374354 and No. 11174360, and the Strategic Priority Research Program (B) of the Chinese Academy of Sciences  (No. XDB07020000).
\end{acknowledgments}


\end{document}